\documentclass[12pt]{article}
\usepackage{amssymb}

\begin{document}
\newcommand{\Tr}{\mathop{\rm Tr}\nolimits}
\newcommand{\re}{\mathop{\rm Re}\nolimits}
\newcommand{\im}{\mathop{\rm Im}\nolimits}
\newcommand{\asympt}{\mathop{\sim}}
\newcommand{\leftpartial}{\mathop{\!\stackrel{\leftarrow}{\partial}}\nolimits}
\newcommand{\rightpartial}{\mathop{\!\stackrel{\rightarrow}{\partial}}\nolimits}
\newcommand{\leftD}{\mathop{\stackrel{\leftarrow}{D}}\nolimits}
\newcommand{\rightD}{\mathop{\stackrel{\rightarrow}{D}}\nolimits}
\def\bra#1{\langle #1 |}
\def\ket#1{|#1 \rangle}
\def\aver#1{\langle\, #1 \,\rangle}
\def\slash#1{\not\!#1}
\def\rdslash{\not\!\partial}
\def\ldslash{\not\!\leftpartial}

\let\eps = \varepsilon
\def \be {\begin{equation}}
\def \ee {\end{equation}}
\def \bea {\begin{eqnarray}}
\def \eea {\end{eqnarray}}

\def \nn {{\mathbb N}}
\def \zz {{\mathbb Z}}
\def \cc {{\mathbb C}}
\def \rr {{\mathbb R}}

\def \ii {{\cal I}}
\def \kk {{\cal K}}
\def \aa {{\cal A}}

\def \nc {noncommutative }
\def \ncg {noncommutative geometry }
\def \sf  {string field }
\def \sft {string field theory }
\def \da {\dagger}

\begin{flushright}
SISSA 109/2000/EP \\
{\tt hep-th/0011238}
\end{flushright}

\vskip 2cm

\begin{center} 
{\LARGE Constraints on the tachyon condensate from anomalous symmetries\\}
\vskip 1cm
{\Large Martin Schnabl}\footnote{E-mail: schnabl@sissa.it} \\ 
\vskip 0.5cm
{\it 
Scuola Internazionale Superiore di Studi Avanzati,\\
Via Beirut 4, 34014 Trieste, Italy
and INFN, Sezione di Trieste}
\end{center}

\begin{abstract}
Using anomalous symmetries of the cubic string field theory vertex we derive set of 
relations between the coefficients of the tachyon condensate. They are in agreement 
with the results obtained from level truncation approximation.  
\end{abstract}

\baselineskip=18pt

\section{Introduction}
Since the original formulation of the Sen's conjecture \cite{Sen:Descent} 
there has been significant progress in understanding the nonperturbative aspects of the \sf theory.
Initially the existence of translationally invariant vacuum with conjectured energy density  
was established numerically \cite{SZ,MT}  to a rather high accuracy 
by the level expansion method \cite{Kostelecky} in the Witten's cubic \sf theory 
\cite{Witten:NCGSFT,GrossJevicki,Samuel:1986}. More recently the Sen's conjecture 
has been proved rigorously in the framework of background 
independent \sft \cite{bisft1,bisft2,bisft3}. Nevertheless it seems worth 
continuing to look for the exact tachyon condensate in the original cubic 
\sft since it can teach us many things \cite{Zwiebach}.

Various insights into the nature of the tachyon condensate has already been obtained in 
\cite{Sen:Universality,Rastelli,KosteleckyPotting,Hata,Zwiebach}. One particular suggestion 
for the exact form of the condensate based 
on the \nc geometric nature of the \sf action was made in \cite{MySFT,HKL}.
Independently of this recent suggestion we will show in this letter, that we can obtain 
some new exact information about the tachyon condensate.
We point out that some anomalous symmetries \cite{GrossJevicki,Romans,Rastelli} 
of the \sft vertex can be used to derive an infinite set of identities for any \sf 
which solves the equations of motion. We will see that at level $n$ we get $n$ additional
constraints on the coefficients of the string field which are in reasonable
agreement with the explicit results from level truncation scheme \cite{SZ,MT}.
It would be very interesting if one could find even further symmetries which would
then fix all the coefficients completely.

\section{Anomalous symmetries}

The \sft action as given by \cite{Witten:NCGSFT,GrossJevicki,Samuel:1986} takes 
the form of \nc Chern-Simons action
\begin{equation}\label{action}
S[\Psi]=-\frac{1}{\alpha' g_o^2} \left( \frac 12 \aver{\Psi, Q \Psi} + 
\frac 13 \aver{\Psi,\Psi*\Psi} \right)
\end{equation}
with the \nc multiplication defined by
\begin{equation}
\Psi_1 * \Psi_2 = bpz \left( \bra{V} \Psi_1 \otimes \Psi_2 \right),
\end{equation}
where $bpz$ denotes the $bpz$ conjugation in conformal field theory and the vertex $\bra{V}$ was
reviewed in the oscillator formulation \cite{Taylor,MySFT} and studied in a background independent 
manner in \cite{Rastelli}.

From \cite{GrossJevicki,Romans,Rastelli} we know that the vertex $\bra{V}$ satisfies 
certain identities. 
For us will be important in particular the following ones for $n$ even
\bea\label{ansyms} 
\bra{V} \sum_{i=1}^3 ( L_{-n}^{(i)} - L_{n}^{(i)} ) &=& 3 k_n^x \bra{V},
\nonumber\\
\bra{V} \sum_{i=1}^3 ( J_{-n}^{(i)} + J_{n}^{(i)} ) &=& 3 ( h_n^{gh} + 3\delta_{n,0}) \bra{V},
\eea
where $L_n$ and $J_n$ denote matter Virasoro and ghost current generators respectively.
The constants $k_n^x$ and $h_n^{gh}$ take for $n$ even the following values
\bea
k_n^x &=& \frac{13 \cdot 5}{27} \cdot \frac{n}{2} (-1)^{\frac n2},
\nonumber\\
h_n^{gh} &=& - (-1)^{\frac n2}.
\eea
For $n$ odd there would be extra signs between the generators in (\ref{ansyms}) and the right 
hand side would vanish. We are not interested in this case since it will not lead to
any information about the tachyon condensate. 
Note that the additional term on the right hand side of the second 
equation in (\ref{ansyms}) accounts for the nontensor character of the ghost number current.

Let us study now the variation of the action (\ref{action}) under the infinitesimal
variations of the \sf
\bea
\delta \Psi &=& ( L_{-n} - L_{n} -  k_n^x ) \Psi,
\nonumber\\
\delta \Psi &=& ( J_{-n} + J_{n} -  h_n^{gh} - 3\delta_{n,0} ) \Psi
\eea
respectively.
Under these variations the cubic term in the action is obviously invariant due to the 
invariance of the vertex (\ref{ansyms}). On the other hand we know that the total action 
should also be invariant as long as $\Psi$ satisfies equations of motion. 
Combining these two facts we get from the kinetic term
\bea\label{anids}
\bra{\Psi} [Q, L_n] \ket{\Psi} &=&   -k_n^x \bra{\Psi} Q \ket{\Psi},
\nonumber\\
\bra{\Psi} [Q, J_n] \ket{\Psi} &=&   h_n^{gh} \bra{\Psi} Q \ket{\Psi}.
\eea  
Let us note that both commutators on the left hand side are modes of conformal primary 
fields, the latter being minus the BRST current $J^B$.

\section{Explicit checks} 

To compare the above formulas with the results obtained in level expansion scheme in \cite{SZ,MT} 
one should first of all impose the Siegel gauge condition $b_0 \ket{\Psi}=0$ on the \sf
and simplify the commutators.
For the first equation of (\ref{anids}) one has simply
\be
[Q, L_n] = -n c_0 L_n + \cdots
\ee
where the dots stand for terms which do not contribute. For the second equation
one can use a little trick. Write the left hand side as
\be
\bra{\Psi} [Q, J_n] \ket{\Psi} = - \bra{\Psi} \{J_n^B, b_0 \} c_0 \ket{\Psi}
\ee
where we used the facts that $[Q,J_n]=-J_n^B$ and $b_0 \ket{\Psi}=0$. The anticommutator
can be easily evaluated using the operator product expansion (see e.g. \cite{Polchinski}).
Both formulas (\ref{anids}) thus simplify in the Siegel gauge to
\bea\label{anidsSiegel}
\bra{\Psi} c_0 L_n \ket{\Psi} &=&  \frac 1n k_n^x \bra{\Psi} c_0 L_0^{tot} \ket{\Psi},
\nonumber\\
\bra{\Psi} c_0 (n J_n + L_n^{tot}) \ket{\Psi} &=&   -h_n^{gh} \bra{\Psi} c_0 L_0^{tot} \ket{\Psi},
\eea  
where $ L_n^{tot}$ denotes the total Virasoro generator. These identities 
can be easily checked for the numerical values obtained in \cite{SZ,MT}.
Let us define $r_n^{L,J}$ to be the ratio of the left and right hand sides of the first or 
second equation of (\ref{anidsSiegel}) respectively. Then inserting for simplicity 
the values for the \sf coefficients 
from \cite{SZ} obtained at the level (4,8) we get the following results
\begin{eqnarray*}
r_2^L &=& 1.069,  \qquad r_4^L =1.044, \\
r_2^J &=& 1.004,  \qquad r_4^J =0.939. 
\end{eqnarray*}

We see that the above identities are preserved within $7 \%$. This can be compared 
with the value of the potential which is for the same values about $1.4 \%$ away from the 
expected value. This discrepancy in the errors by a factor of five does not 
necessarily mean 
that there are mistakes neither in the derivation nor in the numerical evaluation. 
In fact we know that the convergence properties of the level truncation approximation 
depends rather strongly on what kind of calculation we are doing. In an unpublished
work we have studied the properties of the \sf algebra unity $\ket{I}$ in the level 
truncation using the universal recursive methods of \cite{Rastelli}. 
Keeping only terms up to level 8 in the unity $\ket{I}$ and during the whole 
calculation we got for example
\bea
L_{-2} \ket{0} * \ket{I} &=& 0.990 L_{-2} \ket{0} + 0.108 L_{-2}^{tot} \ket{0}
- 0.196 L_{-2}^{tot} L_{-2}^{tot} \ket{0} + \cdots,
\nonumber\\
L_{-2}^{tot} \ket{0} * \ket{I} &=& 0.990 L_{-2}^{tot}\ket{0} + 
0.009 L_{-2}^{tot} L_{-2}^{tot} \ket{0} + \cdots,
\eea
where the dots stand for terms which are relatively smaller or of higher levels where
one can understand bigger errors. Looking at these values one might wonder whether 
after all the string algebra unity is unity also for the state $L_{-2} \ket{0}$. 
The experience from calculations at lower levels where the errors are much bigger
suggests that it really converges, hopefully to the correct state. The fact 
that calculations involving matter Virasoro generators converge much more slowly
can be easily traced back to the presence of the Virasoro anomaly.

\end{document}